\documentclass[12pt]{JHEP3}

\pdfoutput=1
\usepackage{bm}
\usepackage{epsfig}
\usepackage{citesort}
\usepackage{graphicx}
\usepackage{multirow}

\def\be{\begin{equation}}
  \def\ee{\end{equation}}
\def\bea{\begin{eqnarray}}
  \def\eea{\end{eqnarray}}

\title{Crossing Statistic: Bayesian interpretation, model selection and resolving dark energy parametrization problem} 

\author{Arman Shafieloo\\
	Institute for the Early Universe, Ewha Womans University\\ 
	Seoul, 120-750, South Korea\\
	E-mail: \email{arman@ewha.ac.kr}}

\keywords{Supernovae, dark energy, cosmological parameter estimation} 
     
\abstract{By introducing Crossing functions and hyper-parameters I show that the Bayesian interpretation of the Crossing Statistics~\cite{Crossing} can be used trivially for the purpose of model selection among cosmological models. In this approach to falsify a cosmological model there is no need to compare it with other models or assume any particular form of parametrization for the cosmological quantities like luminosity distance, Hubble parameter or equation of state of dark energy. Instead, hyper-parameters of Crossing functions perform as discriminators between correct and wrong models. Using this approach one can falsify any assumed cosmological model without putting priors on the underlying actual model
of the universe and its parameters, hence the issue of dark energy parametrization is resolved. It will be also shown that the sensitivity of the method to the intrinsic dispersion of the data is small that is another important characteristic of the method in testing cosmological models dealing with data with high uncertainties. 

}

\begin{document}

\section{Introduction}                        
\label{sec:introduction}

One of the main goals of physical cosmology is to reconstruct the expansion history of the universe and finding the actual model of dark energy. There have been many approaches introduced in the cosmological community to reconstruct the expansion history of the universe and the properties of dark energy. One can categorize them in general to parametric and non-parametric methods~\cite{review_DE}. Non-parametric methods are least biased by initial assumptions but in many cases they are not efficient and estimation of the errors can be also problematic~\cite{daly03,huterer03,wang04,wang05,shafieloo06,shafieloo07,sahni08,zunckel08,shafieloo09,clarkson10,shafieloo10,nesseris10,holsclaw10,crittenden11}. Contrary to non-parametric methods, parametric methods are easy to use and they may seem so efficient in putting tight constraints on the cosmological quantities but if the assumed parameterization being wrong, which can be easily the case since the actual model of the universe is not yet known, then they can be completely misleading and result to wrong conclusions. See~\cite{davis07,serra07,eric08,sollerman09,kilbinger10} for details of data analysis and methods of parametric reconstruction of the properties of dark energy using supernovae data. \\\\

In this paper I introduce Crossing functions and hyper-parameters, motivated by the idea of the Crossing Statistic~\cite{Crossing}, as a tool to discriminate between correct and wrong models. It is known that if an assumed model is indeed the correct one to describe Gaussian distributed data, then the histogram of the normalized residuals should also have a Gaussian distribution, with zero mean and a standard deviation of 1. If the histogram instead exhibits significant deviation from the a Gaussian distribution, however, then this can be used to rule out the assumed model. The Crossing Statistic~\cite{Crossing} pushes this well known idea from statistical analysis a step further by pointing to the fact that not only should the whole sample of residuals have a Gaussian distribution around the mean, but so should any continuous sub-sample. In other words we have shown that catching the systematic correlations in the residuals can be used to falsify assumed models. In the current paper I introduce the Bayesian interpretation of the Crossing Statistic looking for the systematic correlations in the residuals by comparing an assumed model with its own possible variations. In this approach observable quantities of any assumed model is multiplied by a Crossing function and then is fitted to the data and the resultant values of the hyper-parameters of the Crossing function decide about consistency of the model and the data. In fact derived values of the Crossing hyper-parameters indicate that how smooth deviations from the assumed model can improve the fit to the data. If improvement in the fit to the data being significant by smooth deviation from the assumed model, it hints towards the inconsistency of the model and the data. All these can be done in the framework of $\chi^2$ statistics to define the confidence limits.  Using supernovae type Ia simulated data I will show how efficiently Crossing Statistics can discriminate between cosmological models while by using the standard statistics we need to assume a parametric form for the cosmological quantities. The core idea behind this work is to suggest that instead of parametrizing cosmological quantities like luminosity distance, Hubble parameter or equation of state of dark energy, we can simply falsify any assumed cosmological model by parameterizing the possible deviations from its actual predictions for the cosmological observables. \\\\

In the following I will introduce the Crossing functions and their hyper-parameters and argue how they can be used as discriminators between cosmological models. I will then use simulated data to show how effectively Crossing Statistics can distinguish among cosmological models. Minimal sensitivity of the method to the intrinsic dispersion of the data will be discussed later before I summarize the results at the end.

\section{Method and Analysis}                        
\label{meth}


If the real Universe differs from the assumed theoretical model, one
would hope that it would be possible to develop a statistical test
that would be able to pick up on this.  To these ends we consider the
`crossings' between the predictions of a given model, and the real
Universe from which data is derived. In Crossing Statistic~\cite{Crossing} the existence of these type of crossings were used to develop a new statistic to determine the goodness of fit between an assumed model and the real Universe.

To apply Crossing Statistic on the supernovae data we must first pick a theoretical or phenomenological model of dark energy (e.g. $\Lambda$CDM) and a data set of SN Ia distance moduli $\mu_i(z_i)$. We use the $\chi^2$ statistic to find the best fit form of the assumed model, and from this we then construct the {\it error normalized difference} of the data from the best fit distance modulus $\bar \mu(z)$:
\begin{equation} 
q_i(z_i)=\frac{\mu_i(z_i) - {\bar \mu(z_i)}}{\sigma_i (z_i)}. \label{qidef}
\end{equation}

In {\it one-point Crossing Statistic}, which  
tests for a model and a data set that cross each other at only one
point, we first find this crossing point, which we label by $n_1^{CI}$ and
$z_1^{CI}$.  To achieve this we define
\begin{equation}
 T(n_1)=Q_1(n_1)^2+Q_2(n_1)^2,
\label{qndef} 
\end{equation}   
where $Q_1(n_1)$ and  $Q_2(n_1)$ are given by
\begin{eqnarray}
 Q_1(n_1)&=&\sum_{i=1}^{n_1} q_i(z_i)    \nonumber \\
 Q_2(n_1)&=&\sum_{i=n_1+1}^N q_i(z_i),
\end{eqnarray}
and $N$ is the total number of data points.
If $n_1$ is allowed to take any value from $1$ to $N$ (when the data is sorted by red shift)
then we can maximize $T(n_1)$ to find $T_{I}$ by varying with respect to $n_1^{CI}$. Finally, we use Monte Carlo simulations to find how often we
should expect to obtain a $T_{I}$ larger or equal to the value derived from the
observed data. In doing this we find the fraction of
Monte Carlo data sets leading to $T_{I}^{M.C} \geq T_{I}^{data}$,
which we will use as an estimate of the probability that the data set should
be realized from the particular best fit cosmological model we have
been considering. 

This approach can be extended to models with more than one crossing
point by the {\it two-point Crossing Statistic}. In this case we
assume that the model and the data cross each other at two points and,
as above, we try to find the two crossing points and their red shifts,
which we now label $n_1^{CII},z_1^{CII}$ and $n_2^{CII},z_2^{CII}$.
This is achieved by defining
\begin{equation}
 T(n_1,n_2)=Q_1(n_1,n_2)^2+Q_2(n_1,n_2)^2+Q_3(n_1,n_2)^2,
 \end{equation}   
where the $Q_i(n_1,n_2)$ are now given by
\begin{eqnarray}
 Q_1(n_1,n_2) &=&\sum_{i=1}^{n_1} q_i(z_i) \nonumber \\
 Q_2(n_1,n_2) &=&\sum_{i=n_1+1}^{n_2} q_i(z_i) \nonumber \\
 Q_3(n_1,n_2) &=&\sum_{i=n_2+1}^{N} q_i(z_i).
 \label{eq:QT2} 
\end{eqnarray}
We can then maximize $T(n_1,n_2)$ by varying with respect to $n_1$
and $n_2$, to get $T_{II}$.  Comparing $T_{II}$ with the results from
Monte Carlo realizations then allows us to determine how often
we should expect a two-point crossing statistic that is greater than
or equal to the $T_{II}$ obtained from real data.  The {\it three-point
  Crossing Statistic}, and higher statistics, can be defined in a
similar manner.  This can continue up to the {\it N-point Crossing
  Statistic} which is, in fact, identical to $\chi^2$~\cite{Crossing}.  We
also note that the zero-point Crossing Statistic, $T_{0}=(\sum_{i}^N q_i)^2$, 
is very similar to the Median Statistic developed by Gott {\it et al.}~\cite{gott01}.  The Crossing Statistic can therefore be thought of as generalizing both the $\chi^2$ and Median Statistics, which it approaches in different limits.

The core idea of Crossing Statistics is based on the fact that any wrong assumed model and the data
cross each other at one or two or maybe even more points within the range of data. While in the previous work that we proposed the Crossing Statistic we had a pure frequentist approach, in the current paper we introduce the Bayesian interpretation of Crossing Statistic which makes it easier to use and apply on the cosmological data.  

In this paper I argue that since within the frame work of FLRW universe the cosmic distances (distance modules) monotonically increases by redshift, hence any two cosmological models can become indistinguishable (up to a very high precision of the data) if we multiply the distance modules of one of them by a function form  with a certain degree and for some particular coefficients. The Bayesian interpretation of Crossing Statistic is in fact hidden in these two prior assumptions that 1) $\mu(z)$ increases by redshift monotonically for all cosmological models hence there are no high frequency fluctuations in $\mu(z)$ and 2) since the distance modules of all cosmological models increases by redshift, $\mu(z)$ of any two cosmological models can become so close to each other at all redshifts up to an indistinguishable level if we multiply $\mu(z)$ of one of them to a suitable smooth function of degree $n$ with some particular values for the coefficients. Coefficients of these functions are in fact the Crossing hyper-parameters that if all of them be simultaneously consistent to zero value then we consider the assumed model to be consistent to the data. 

In Bayes theorem, the posterior $P(M|D)$ that is the probability of a hypothesis (assumed model with some particular parameters) given the data is related to a) prior $P(M)$ that indicates our beliefs about how likely the assumed hypothesis is (independent of the observed data) b) likelihood  $P(D|M)$ that represents the likeliness of the observed data assuming the hypothesis and c) $P(D)$ model evidence that is constant for different hypotheses and does not enter in to determination of the probabilities:

\begin{equation}
P(M|D)=\frac{P(D|M)P(M)}{P(D)}
\end{equation} 

In our analysis we assume that the actual underlying model would be covered by a point in the hyperparameter space of the proposed model (any of these points in the hyperparameter space represent a hypothesis) and all these hypotheses are equally likely before observing the data. So we use a flat prior for all hypotheses in the hyperparameter space and our posterior would become directly proportional to the likelihood. \\ 

Since the behavior of matter density and its effect on the expansion history of the universe is known, for any assumed model of dark energy we keep $\Omega_{0m}$ as a free parameter along with Crossing hyper-parameters of the function form. We multiply the luminosity distance given by our assumed dark energy model $\mu_{DE}(\Omega_{0m},z)$ to the function $F(C_1,...C_N,z)$ and fit this new function to the data using $\chi^2$ minimization and find the best fit point in the hyper-parameter space and derive the confidence limits. \\

Crossing function is a parametrization proposed to pick up the possible deviations from the assumed theoretical model using the data. The basic difference between the Bayesian interpretation of the Crossing statistic and usual parametrizations of cosmological quantities is in the fact that we are not comparing cosmological models within a certain framework of a parametric form and finding the best fit point in the parameter space and defining the confidence limits, but instead we are comparing an assumed cosmological model with its own possible variations. If the data suggest that the deviations from the actual model must be significant we can rule out the model right away. In fact parameterizations of cosmological quantities can never be used to falsify the assumed parametric form and we can only find the best fit point in the parameter space while the Bayesian interpretation of Crossing statistic is designed to falsify any assumed theoretical model. Crossing function can have different forms but considering the shape of the data and our expectations from the theoretical models (once again, $\mu(z)$ should increase smoothly and monotonically by redshift) we expect a simple Chebyshev polynomials \cite{barlow} can perform satisfactorily. Chebyshev polynomials have properties of orthogonality and convergence within the limited range of $-1 < x < 1$ which is appropriate in our case of study. One can use Gram-Schmidt process to generate a set of orthogonal basis tailored appropriately for the case of study but in our analysis and considering the range of the data Chebyshev polynomials are good enough. \\

To falsify a theoretical model one needs to fit the given theoretical model along with the Crossing functions of different degrees. Looking at the data having more than two crossing between the actual model and the assumed model seems to be unrealistic~\cite{Crossing} and even if it is possible for some special models, it would not be detectable by the data since there are uncertainties in the data and if two models be so close to each other, data cannot discriminate between them. However, one needs not to be worry about this since assuming a Crossing function of high degree increases the allowed $\Delta \chi^2$ with respect to the $\chi^2$ of the best fit point for different confidence levels so automatically analysis will show that by increasing the degrees of freedom of the function form the data would not be sensitive to pick up the tiny differences. \\

In this analysis I assume Chebyshev polynomials of orders one and two as the Crossing functions and to show the performance of the method, I do test the spatially flat $\Lambda$CDM model for three sets of simulated data according to three dark energy models (all spatially flat universes), an evolving dark energy model or Kink model with the same parameters used in~\cite{Corasaniti,LANL} \footnote{equation of state of dark energy in this particular model is given by: 

\begin{eqnarray}
\label{eq:model3}
&&w(z) =
w_0+(w_m-w_0)\frac{1+\exp(\Delta_t^{-1}(1+z_t)^{-1})}{1-\exp(\Delta_t^{-1})}\\ 
&&\times\left[ 1-
\frac
{\exp(\Delta_t^{-1}) + \exp(\Delta_t^{-1}(1+z_t)^{-1})}
{\exp(\Delta_t^{-1}(1+z)^{-1}) +\exp(\Delta_t^{-1}(1+z_t)^{-1})} 
\right],\nonumber
\end{eqnarray}
with the constants having the values $w_0 = -1.0, ~w_m = -0.5,~z_t =
0.5,~\Delta_t = 0.05$. } with $\Omega_m=0.27$, a model of dark energy with constant equation of state of $w(z)=0.9$ with $\Omega_m=0.27$ and finally $\Lambda$CDM model with $w(z)=-1$ and $\Omega_m=0.27$ (correct model). Simulated data is based on future JDEM data with 2298 data points in the range of $0.015<z<1.7$~\cite{JDEM}. The Kink model I used in this analysis has a special form of the equation of state of dark energy that at low redshifts it converges to $w(z)=-1$ and at higher redshifts it smoothly changes to $w=-0.5$. This evolving equation of state of dark energy is not within the possibilities of CPL~\cite{CPL} or many other dark energy parametrizations. Hence using these parameterizations results to wrong reconstruction of dark energy if the data is based on the Kink model (look at upper-right panel of figure~6 in~\cite{LANL}). 

Crossing functions are given by:

\begin{equation}
F_{I}(C_1,z)=1+C_1(\frac{z}{z_{max}})
\end{equation}

and

\begin{equation}
F_{II}(C_1,C_2,z)=1+C_1(\frac{z}{z_{max}})+C_2[2(\frac{z}{z_{max}})^2-1],
\end{equation}

where $z_{max}$ is the maximum redshift in the data set. The functions that we fit to the data are $\mu_{DE}^{F_I}(z)=\mu_{DE}(\Omega_{0m},z) \times F_{I}(C_1,z)$ and $\mu_{DE}^{F_{II}}(z)=\mu_{DE}(\Omega_{0m},z) \times F_{II}(C_1,C_2,z)$ and variables are {$\Omega_{0m}, C_1$} in case of one crossing and {$\Omega_{0m}, C_1, C_2$} in case of two crossings. $\mu_{DE}(\Omega_{0m},z)$ represents the luminosity distance of the assumed dark energy model with matter density of $\Omega_{0m}$. As it was mentioned earlier one can go to higher orders of crossing but we will see that the simulated data we are using are not sensitive to the higher orders. The rest is the usual likelihood analysis based on $\chi^2$ statistics. In~\cite{Crossing} we have explored that $\chi^2$ statistic is in fact equivalent to the last mode of Crossing statistic and in the current analysis in fact we use an approximation to transfer the information of the first and second Crossing modes via the Crossing function to the last mode which is $\chi^2$. One may argue that the form of Crossing function may not be appropriate (to map the wrong model to the actual model) but considering the  form and range of the data Chebyshev polynomials seems to be the most appropriate choice to do this mapping to the level of indistinguishability between the models. 

\begin{figure*}[!t]
\centering
\begin{center}
\vspace{-0.05in}
\centerline{\mbox{\hspace{0.in} \hspace{2.1in}  \hspace{2.1in} }}
$\begin{array}{@{\hspace{-0.3in}}c@{\hspace{0.3in}}c@{\hspace{0.3in}}c}
\multicolumn{1}{l}{\mbox{}} &
\multicolumn{1}{l}{\mbox{}} \\ [-2.8cm]
\hspace{-0.4in}
\includegraphics[scale=0.3, angle=0]{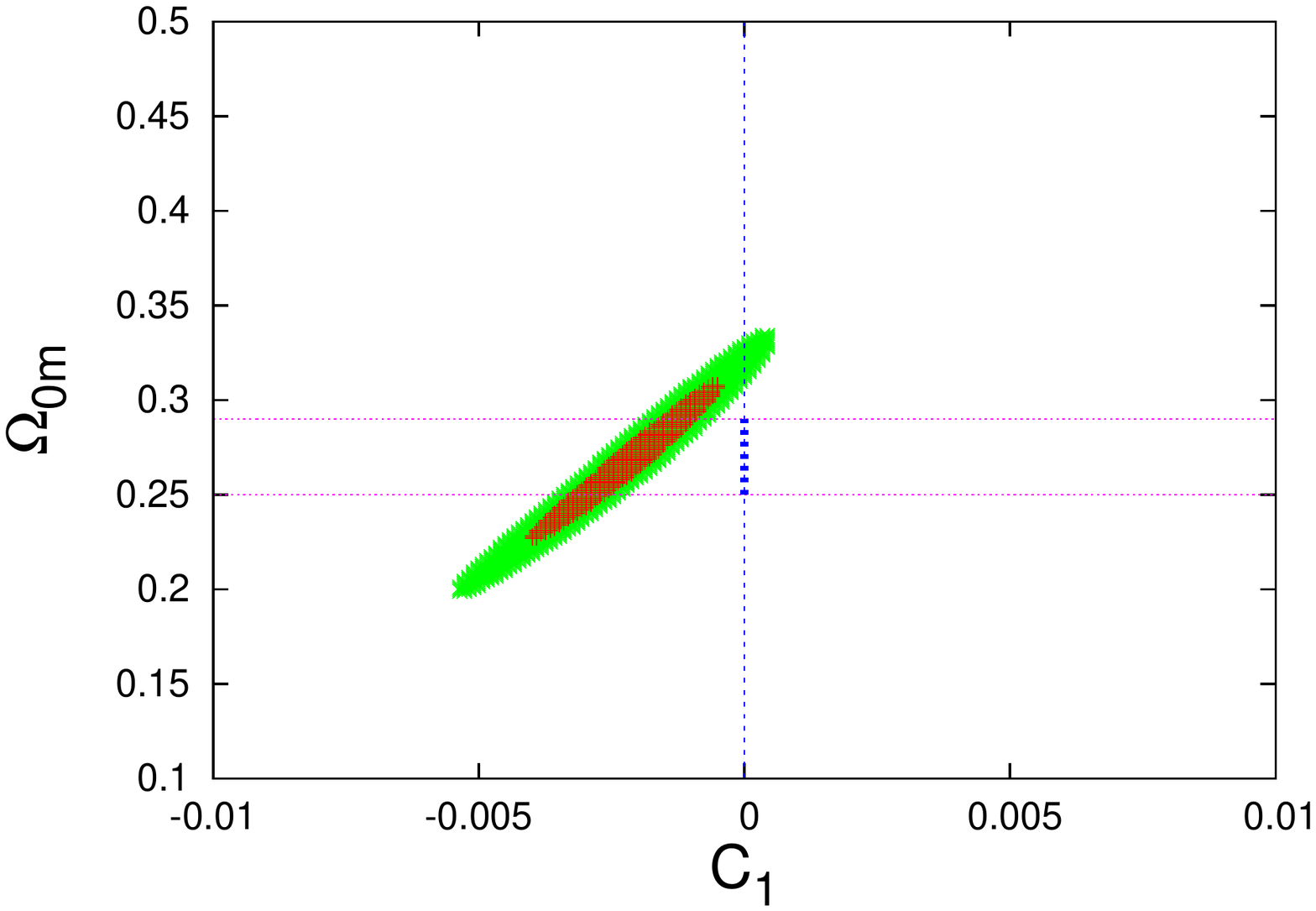}
\hspace{-1.in}
\includegraphics[scale=0.3, angle=0]{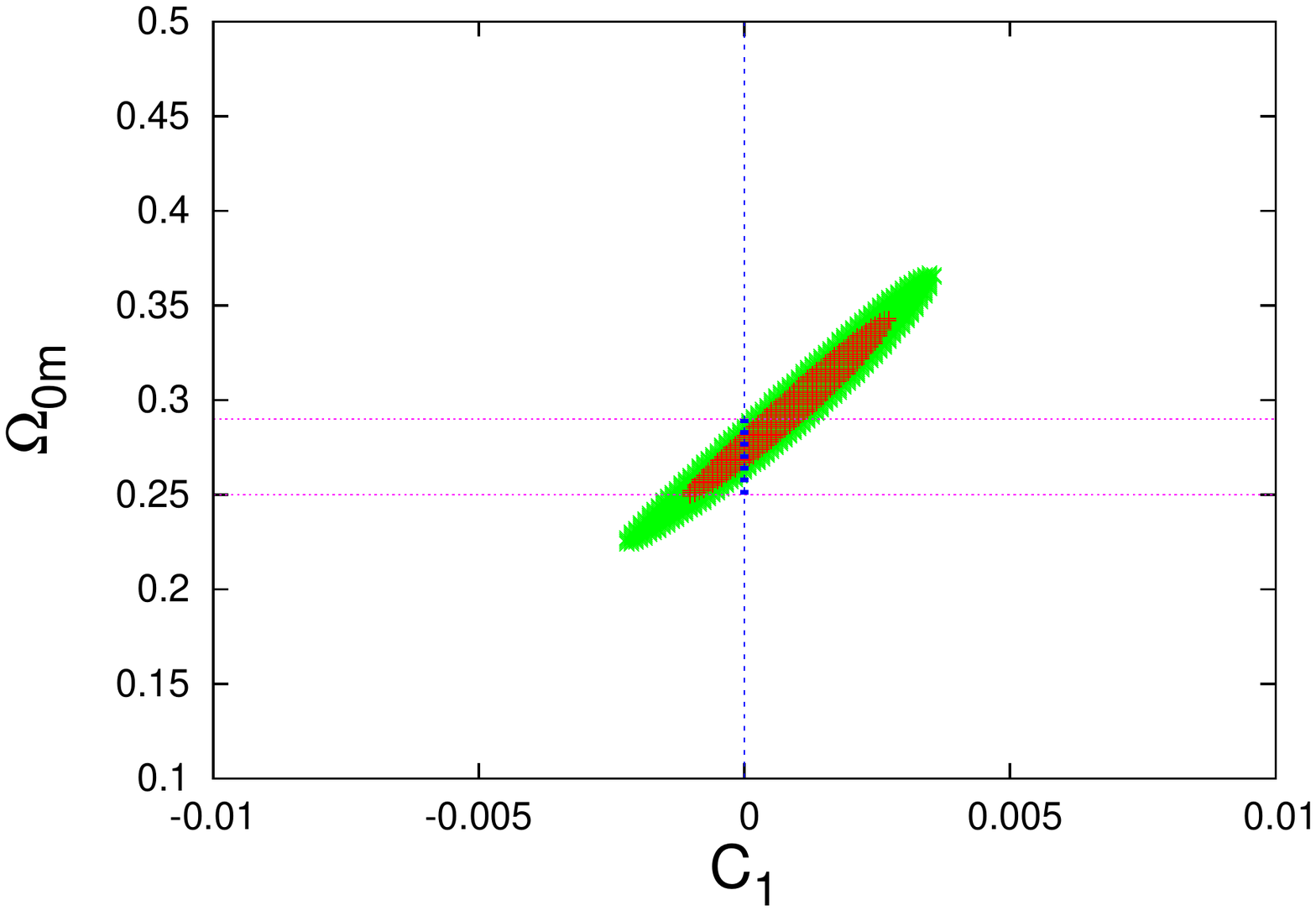}
\hspace{-1.in}
\includegraphics[scale=0.3, angle=0]{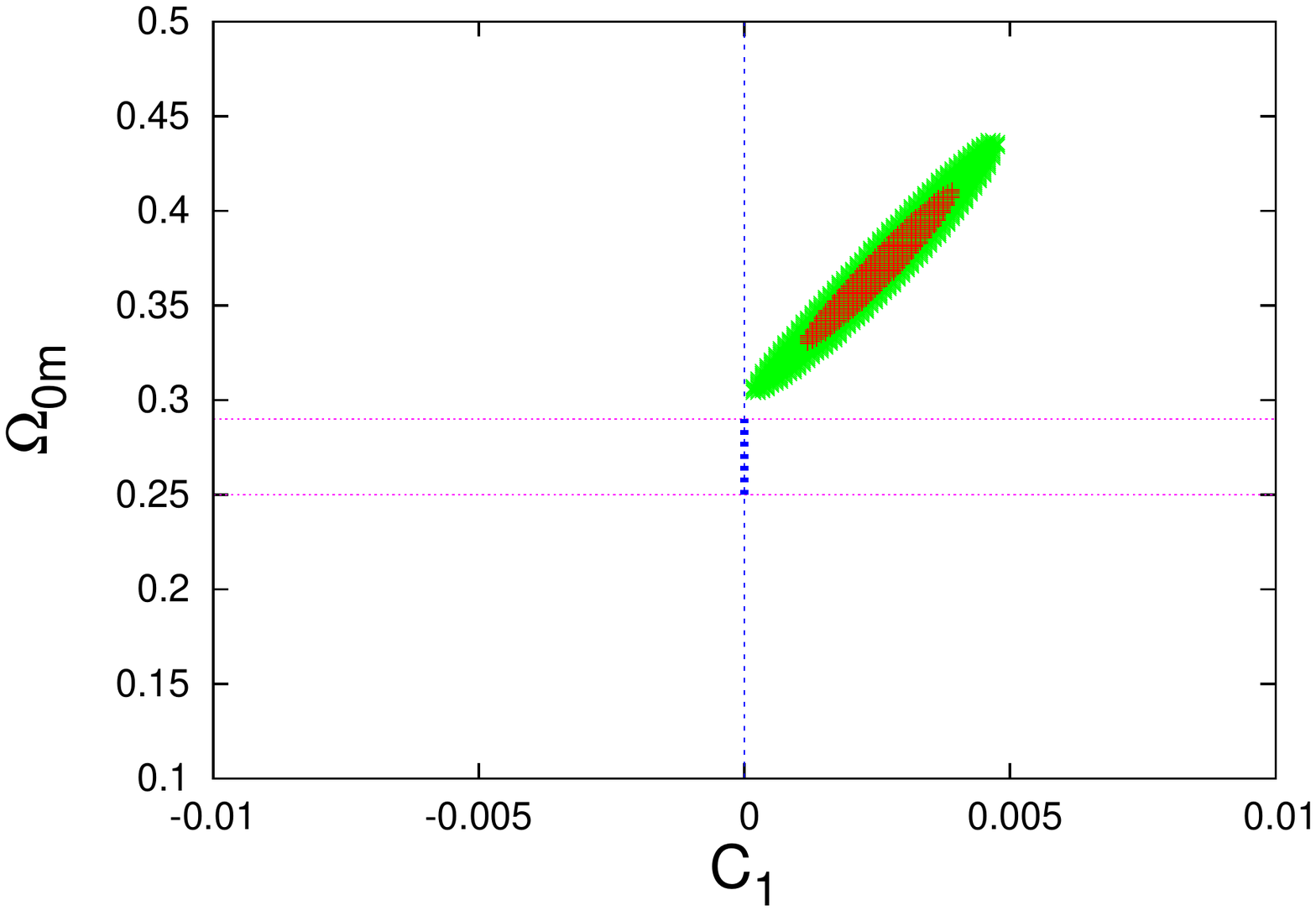}
\hspace{-1.1in}
\vspace{-0.2in}
\end{array}$
\vspace{-0.2in}
\vspace{-0.3in}
\end{center}
\caption {\small $1\sigma$ and $2\sigma$ confidence limits of fitting $\mu_{DE}^{F_I}(z)$ assuming the theoretical model being $\Lambda$CDM to three different simulated data sets. In the left panel the data is generated based on kink dark energy model with $\Omega_{0m}=0.27$, the data in the middle panel is based on the  $\Lambda$CDM model ($w=-1$) with $\Omega_{0m}=0.27$  and in the right panel data is based on quiescence dark energy model with $w=-0.9$ and $\Omega_{0m}=0.27$. The vertical line at $C_1=0$ represents the consistency of the assumed model and the data. Two horizontal lines at $\Omega_{0m}=0.25$ and $\Omega_{0m}=0.29$ represents possible constraints we may be able to put on the matter density from other cosmological observations to help us falsifying cosmological models.}  
\label{fig:T1}
\end{figure*}

In figure.\ref{fig:T1} we see the results of fitting $\mu_{DE}^{F_I}(z)$ assuming the theoretical model being $\Lambda$CDM to three different simulated data sets. The data in the left panel is based on kink dark energy model with $\Omega_{0m}=0.27$, the data in the middle panel is generated based on  $\Lambda$CDM model ($w=-1$) with $\Omega_{0m}=0.27$ and the right panel is for the data based on quiescence dark energy model with $w=-0.9$ and $\Omega_{0m}=0.27$. All models are spatially flat and the data is according to the future JDEM data.

\begin{figure*}[!t]
\centering
\begin{center}
\vspace{-0.65in}
\centerline{\mbox{\hspace{0.in} \hspace{2.1in}  \hspace{2.1in} }}
$\begin{array}{@{\hspace{-0.3in}}c@{\hspace{0.3in}}c@{\hspace{0.3in}}c}
\multicolumn{1}{l}{\mbox{}} &
\multicolumn{1}{l}{\mbox{}} \\ [-2.8cm]
\hspace{0.9in}
\includegraphics[scale=0.5, angle=0]{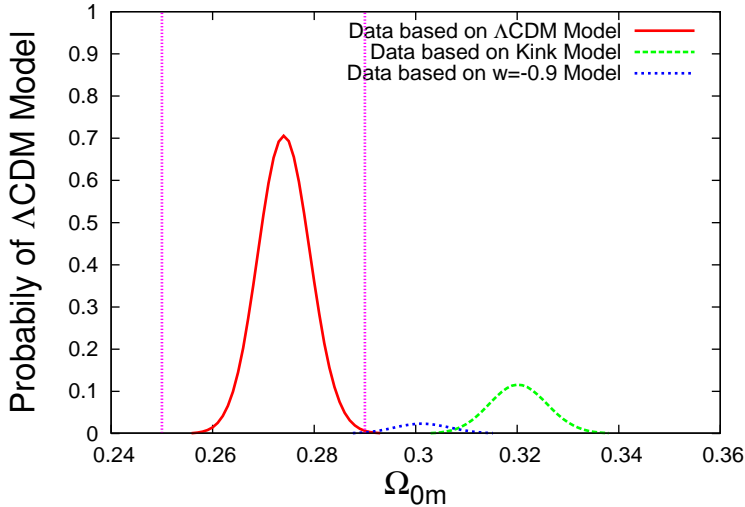}
\vspace{-0.2in}
\end{array}$
\vspace{-0.2in}
\vspace{-0.3in}
\end{center}
\caption {\small Absolute probability distribution function of $\Lambda$CDM model with different values of $\Omega_{0m}$ for three different simulated data sets. The green-dashed line represents the absolute probability distribution function of $\Lambda$CDM model fitting the data based on kink dark energy model with $\Omega_{0m}=0.27$, the red-solid line represents the probability distribution of $\Lambda$CDM model fitting the data based on the  $\Lambda$CDM model ($w=-1$) with $\Omega_{0m}=0.27$ and the blue-dotted line represents the probability distribution of $\Lambda$CDM model fitting the data based on quiescence dark energy model with $w=-0.9$ and $\Omega_{0m}=0.27$. The vertical line at $\Omega_{0m}=0.25$ and $\Omega_{0m}=0.29$ represents possible constraints we may be able to put on the matter density from other cosmological observations to help us falsifying cosmological models.}  
\label{fig:Prob}
\end{figure*}

\begin{figure*}[!t]
\centering
\begin{center}
\vspace{-0.05in}
\centerline{\mbox{\hspace{0.in} \hspace{2.1in}  \hspace{2.1in} }}
$\begin{array}{@{\hspace{-0.3in}}c@{\hspace{0.3in}}c@{\hspace{0.3in}}c}
\multicolumn{1}{l}{\mbox{}} &
\multicolumn{1}{l}{\mbox{}} \\ [-2.8cm]
\hspace{-0.4in}
\includegraphics[scale=0.3, angle=0]{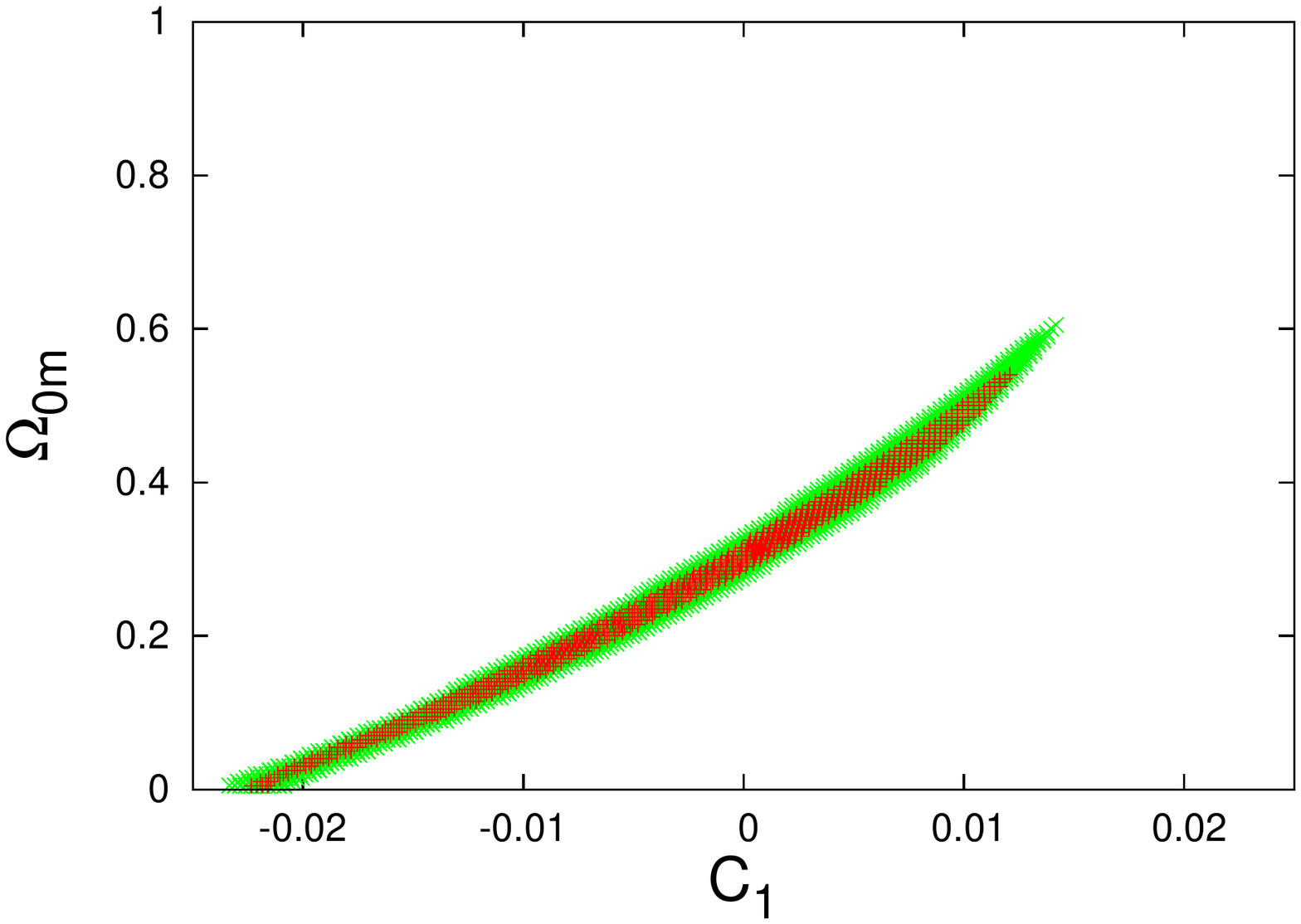}
\hspace{-1.in}
\includegraphics[scale=0.3, angle=0]{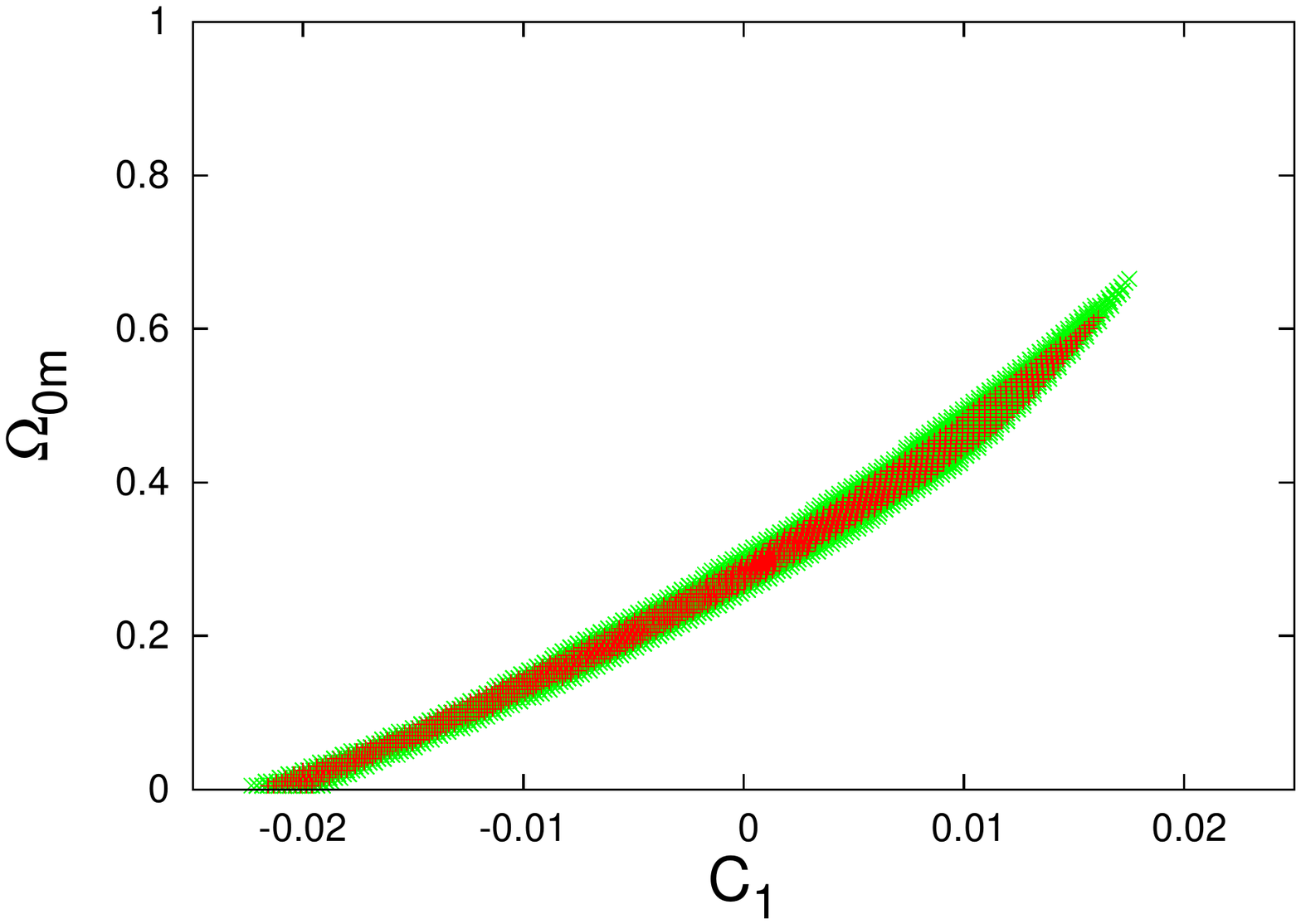}
\hspace{-1.in}
\includegraphics[scale=0.3, angle=0]{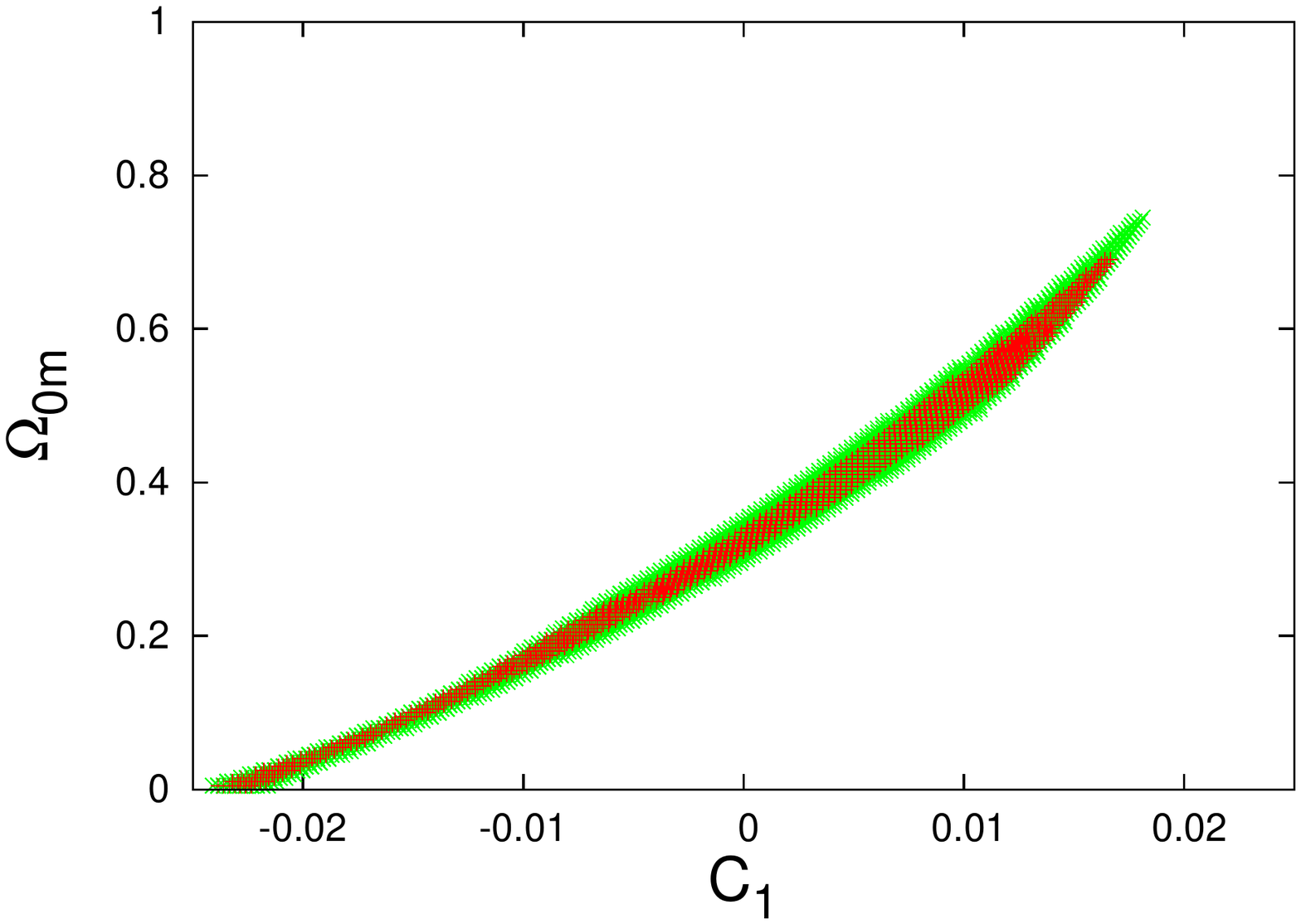}
\hspace{-1.1in}
\vspace{0.in}
\end{array}$
\vspace{0.2in}
$\begin{array}{@{\hspace{-0.3in}}c@{\hspace{0.3in}}c@{\hspace{0.3in}}c}
\multicolumn{1}{l}{\mbox{}} &
\multicolumn{1}{l}{\mbox{}} \\ [-2.8cm]
\hspace{-0.4in}
\includegraphics[scale=0.3, angle=0]{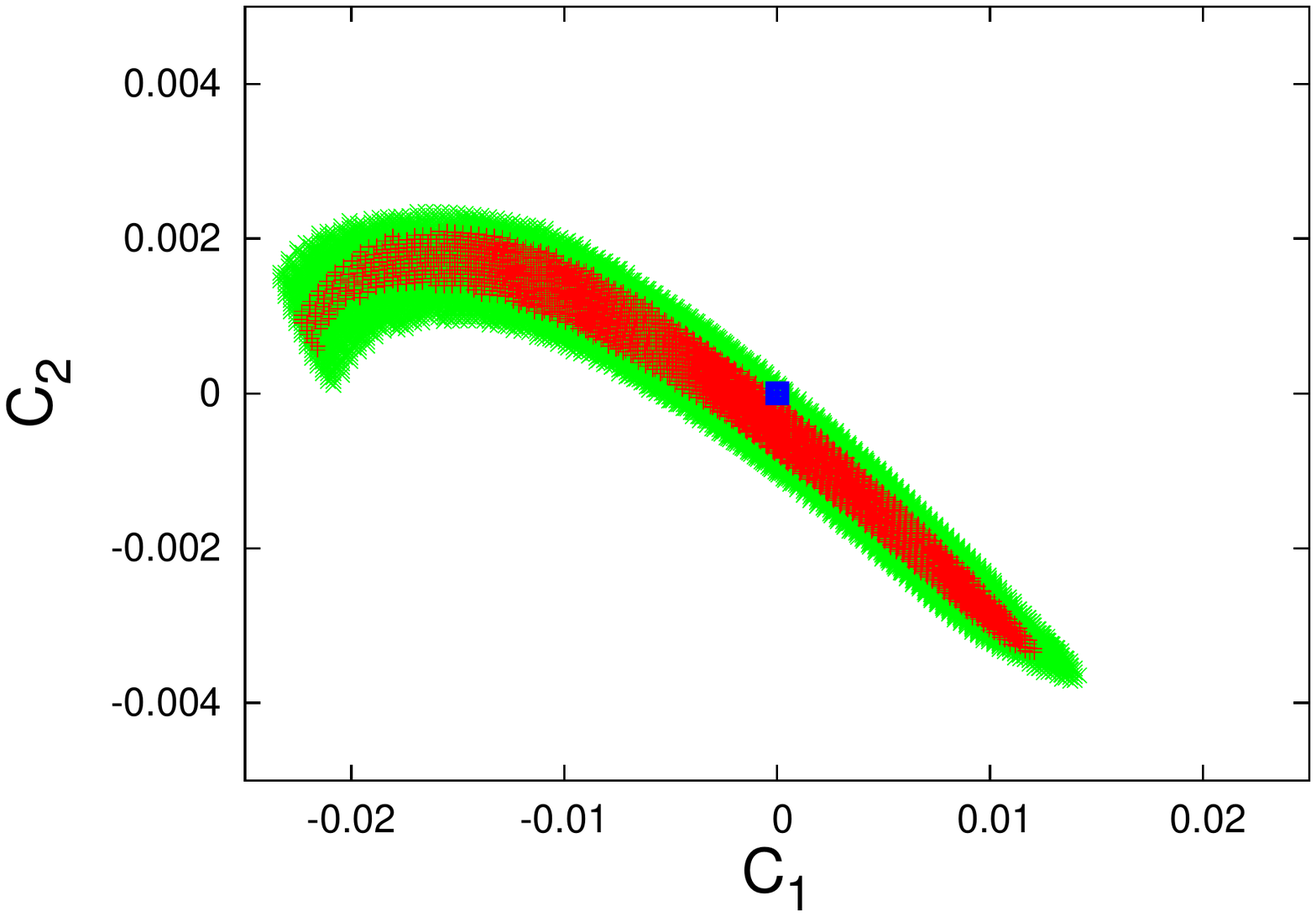}
\hspace{-1.in}
\includegraphics[scale=0.3, angle=0]{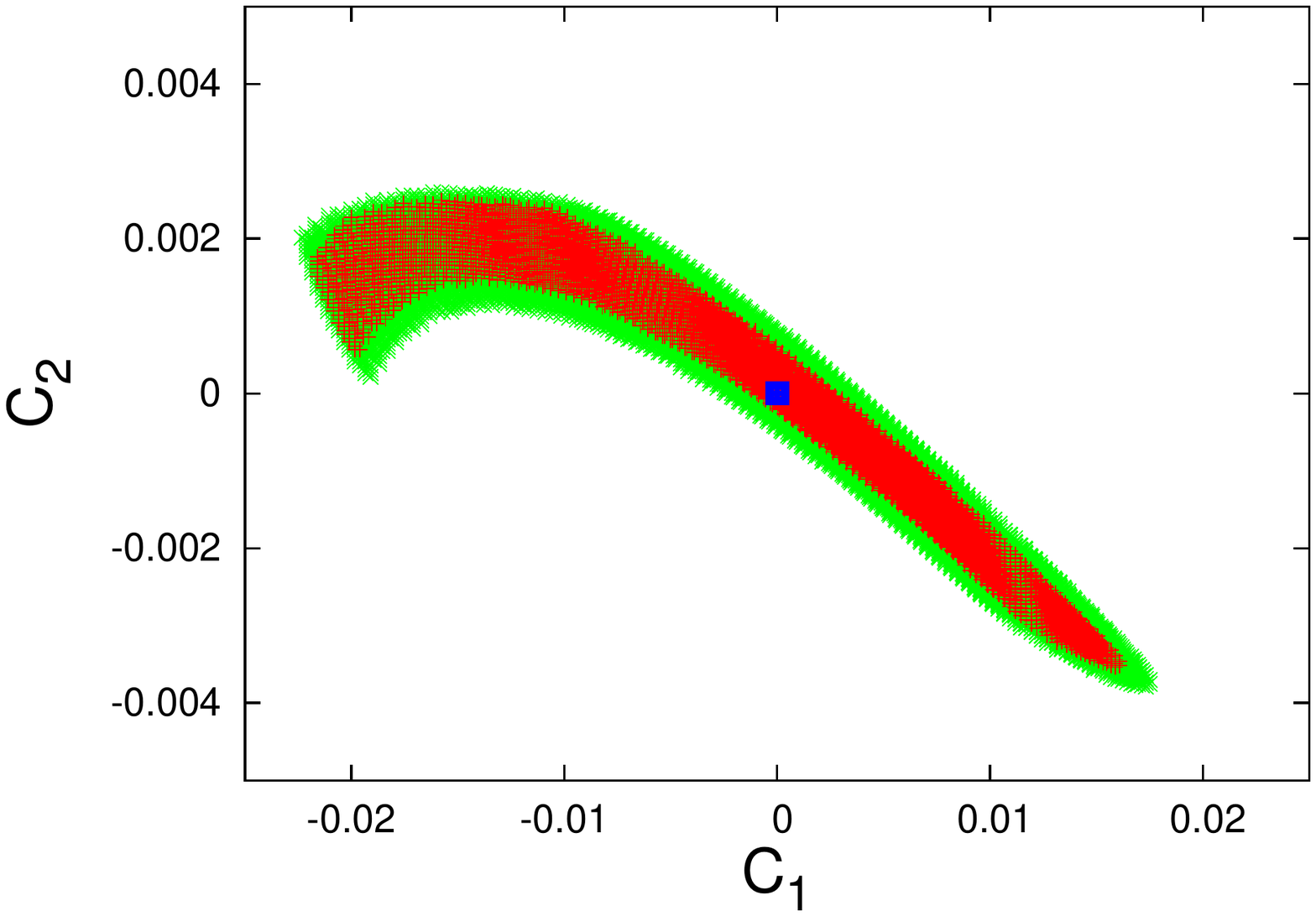}
\hspace{-1.in}
\includegraphics[scale=0.3, angle=0]{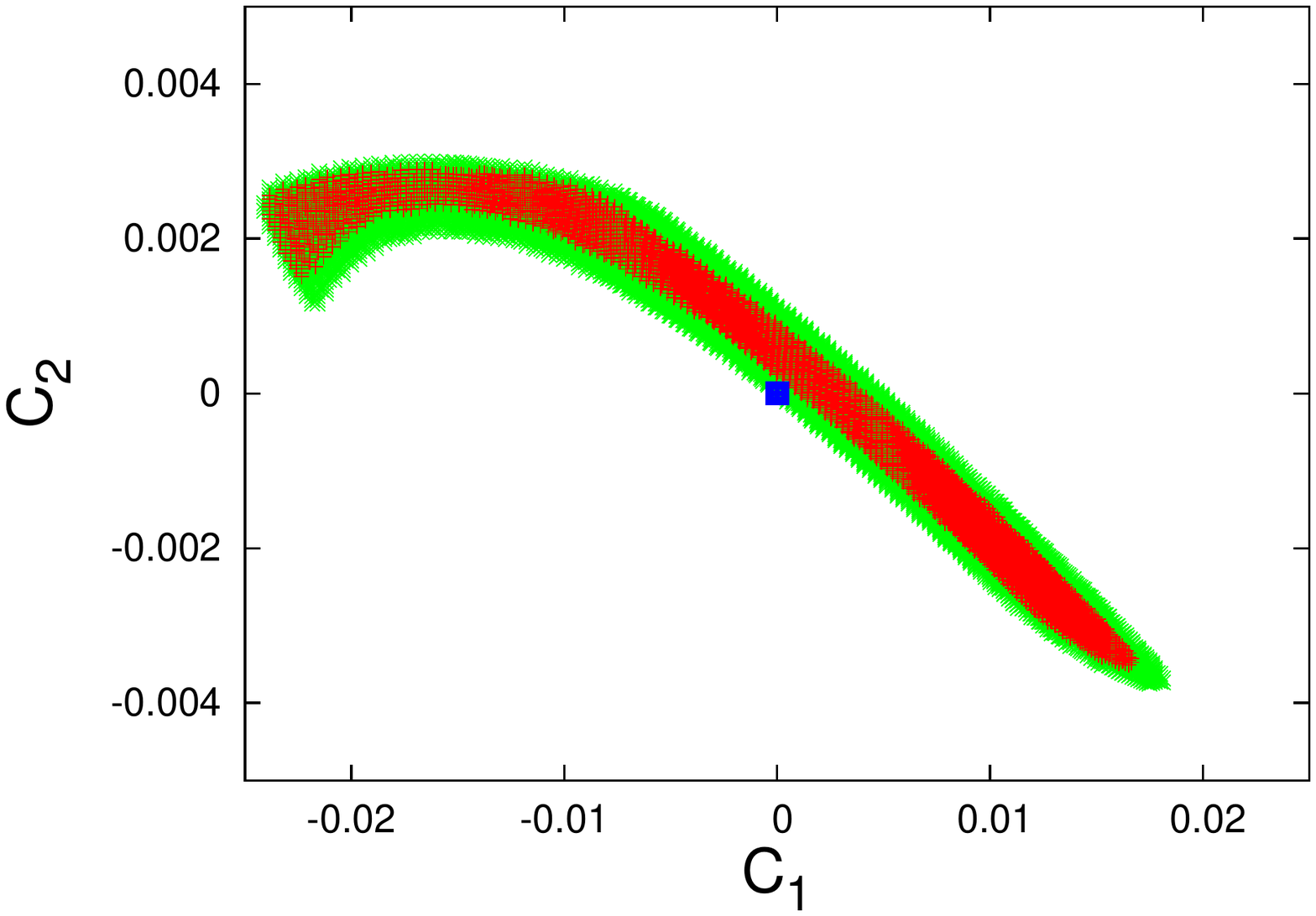}
\hspace{-1.1in}
\vspace{-0.2in}
\end{array}$
\vspace{-0.3in}
\end{center}
\caption {\small $1\sigma$ and $2\sigma$ confidence limits of $C_1$-$\Omega_{0m}$ (upper panels) and $C_1$-$C_2$ (lower panels) of fitting $\mu_{DE}^{F_{II}}(z)$ assuming the theoretical model being $\Lambda$CDM to three different simulated data sets similar to figure 1 (data is based on  kink dark energy model in the left, $\Lambda$CDM in the middle and $w=-0.9$ in the right, all with $\Omega_{0m}=0.27$). The blue dots in the lower panels represents $C_1=C_2=0$ which shows the consistency of the assumed model and the data. One can see that the data are not sensitive to the second order function form $F_{II}$ to distinguish between the cosmological models.}
\label{fig:T2}
\end{figure*}

We can clearly see that the $w=-0.9$ model can be ruled out with high confidence and the kink dark energy model shows inconsistency with the data at one to two sigma with no knowledge of matter density. Having a knowledge of matter density from other cosmological observations we can rule out the kink dark energy model too as it suggest the value of matter density should be about $0.32$ for zero value of $C_1$ for standard flat $\Lambda$CDM model. \\

One of the biggest strengths of the Bayesian approaches in performing data analysis is in the ability of the method to estimate the posterior or the probability of the proposed model given the data. As we have discussed earlier, assuming a flat prior for our hypotheses we can directly associate the posterior to the likelihood of the points in the hyperparameter space and estimate the probability of the assumed models given the data. This probability for a point in the hyperparameter space with $\Delta \chi^2= X$ with respect to the best fit point is given by~\cite{barlow}:

\begin{eqnarray}
\label{eq:prob}
P(X;N)=\frac{2^{-N/2}}{\Gamma(N/2)}X^{\frac{N-2}{2}} \exp(-X/2)\\ \nonumber
Prob(X;N)=\int_X^{\infty}P(X';N)dX'
\end{eqnarray}

where $N$ is the degrees of freedom and $\Gamma$ is Gamma function. In case of assuming one crossing and by fitting $\mu_{DE}^{F_I}(z)$ to the data we have only two degrees of freedom and for $N=2$ we can simply derive $Prob(X;2)=\exp(-X/2)$. Now for the points with $C_1=0$ in the hyperparameter space (where these points represent the assumed model) we can derive the probabilities of the assumed model (e.g. $\Lambda$CDM model in our analysis) for different values of $\Omega_{0m}$. In figure.~\ref{fig:Prob} we can see the absolute probability distribution function of the assumed $\Lambda$CDM model given the three simulated data sets same as the data being used in figure.~\ref{fig:T1}. One should note that any of these probability functions have been derived independently for each data set. This plot clearly shows the strength of the method in estimation of the probability of a model given the data and falsifying cosmological models without comparing models to each other.

\begin{table*}
\begin{tabular}{ccccccccc}
 $$ & Assumed Model: & $\Lambda$CDM & $$\\
\hline
 $$ &$\chi^2(\mu_{DE})$ & \hspace{-10 mm}$\chi^2(\mu_{DE}^{F_I})$ & $\chi^2(\mu_{DE}^{F_{II}})$ \\
\hline
Data based on $\Lambda$CDM Model with $\Omega_{\rm 0m}=0.27$ & $2368.9$ &  \hspace{-10 mm}$2368.2$ & $2367.9$ \\
\hline
Data based on $Kink$ Model with $\Omega_{\rm 0m}=0.27$ & $2372.5$ & \hspace{-10 mm}$2368.7$ & $2368.5$ \\
\hline
Data based on $w=-0.9$ Model with $\Omega_{\rm 0m}=0.27$ & $2375.5$ & \hspace{-10 mm}$2368.9$ & $2367.9$ \\
\hline
\end{tabular}
\caption{\small Derived $\chi^2$ using $\mu_{DE}$,  $\mu_{DE}^{F_{I}}$ and  $\mu_{DE}^{F_{II}}$ fitting the three different simulated data sets. Note the changes in the $\chi^2$ by increasing the degrees of freedom of the functions. Its clear that in all three cases, $\chi^2(\mu_{DE}^{F_I})$ and $\chi^2(\mu_{DE}^{F_{II}})$ are so close to each other that we can conclude the data is not sensitive to the second order of functions.}
\label{table:data}
\end{table*}

In figure~\ref{fig:T2} we see the results of fitting $\mu_{DE}^{F_{II}}(z)$ assuming the theoretical model being $\Lambda$CDM to three different simulated data sets similar to figure~\ref{fig:T1}.

We see that by increasing the degrees of freedom of the Crossing function, $\Lambda$CDM model becoming more consistent to all three sets of the data which shows that the data is not sensitive to the crossing function of second order for these particular models of dark energy. I should emphasize that all of these three assumed dark energy models are very close to each other when we allow matter density to vary~\cite{arman_eric}. For instance $\Lambda$CDM model with $\Omega_{0m}=0.27$ is extremely close to kink dark energy model with $\Omega_{0m}=0.314$. Looking at figure~\ref{fig:T2} also make it clear that there is no need to go to the higher orders of crossing. It basically means that the data cannot pick up two or three crossings of the comparing models that is clearly understandable and expected considering the quality and the range of the data.

\begin{figure}[!t]
\centering
\begin{center}
\vspace{-0.05in}
\centerline{\mbox{\hspace{0.in} \hspace{2.1in}  \hspace{2.1in} }}
$\begin{array}{@{\hspace{-0.3in}}c@{\hspace{0.3in}}c@{\hspace{0.3in}}c}
\multicolumn{1}{l}{\mbox{}} &
\multicolumn{1}{l}{\mbox{}} \\ [-2.8cm]
\hspace{-0.4in}
\includegraphics[scale=0.3, angle=0]{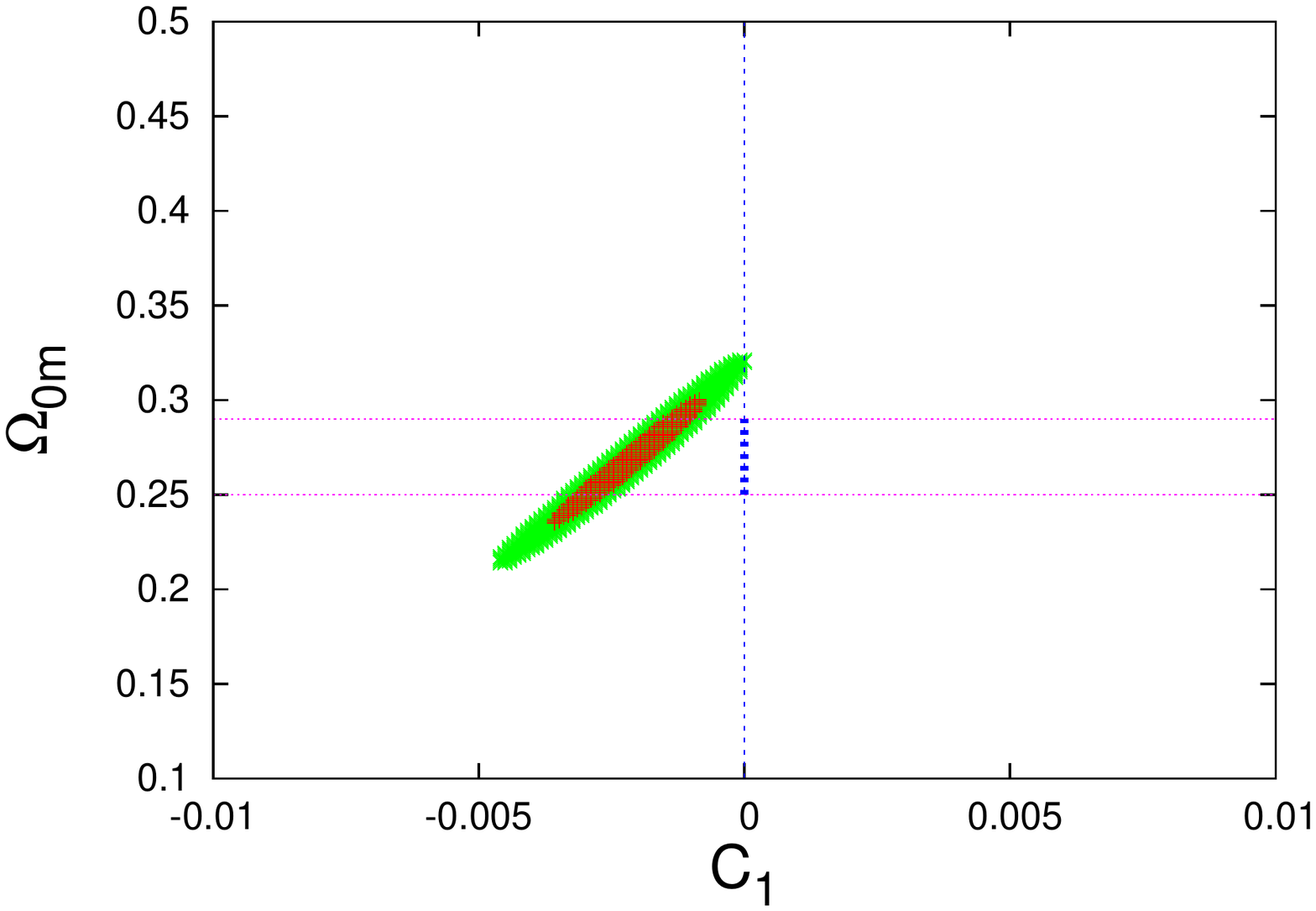}
\hspace{-1.in}
\includegraphics[scale=0.3, angle=0]{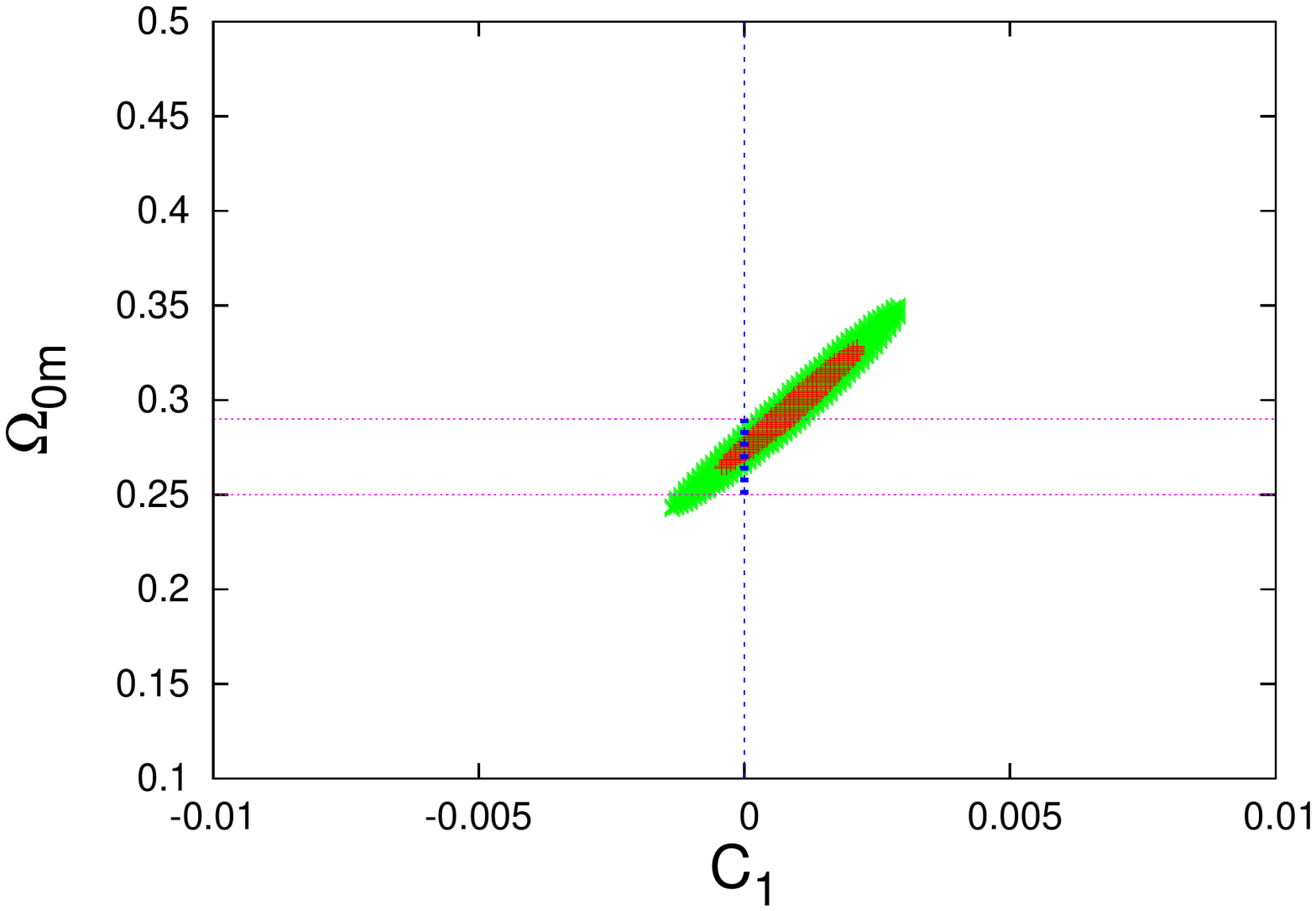}
\hspace{-1.in}
\includegraphics[scale=0.3, angle=0]{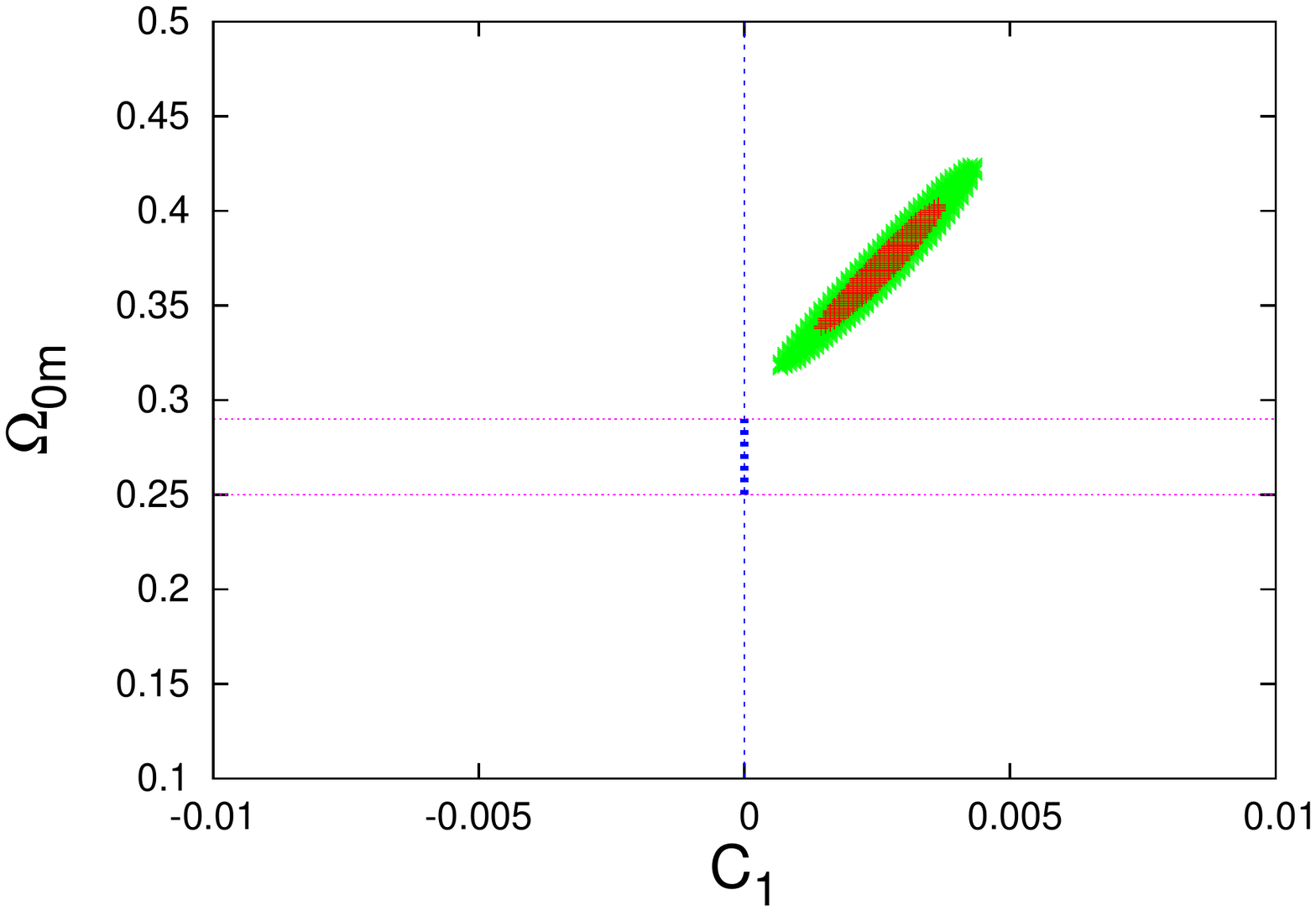}
\hspace{-1.1in}
\vspace{0.in}
\end{array}$
\vspace{0.2in}
$\begin{array}{@{\hspace{-0.3in}}c@{\hspace{0.3in}}c@{\hspace{0.3in}}c}
\multicolumn{1}{l}{\mbox{}} &
\multicolumn{1}{l}{\mbox{}} \\ [-2.8cm]
\hspace{-0.4in}
\includegraphics[scale=0.3, angle=0]{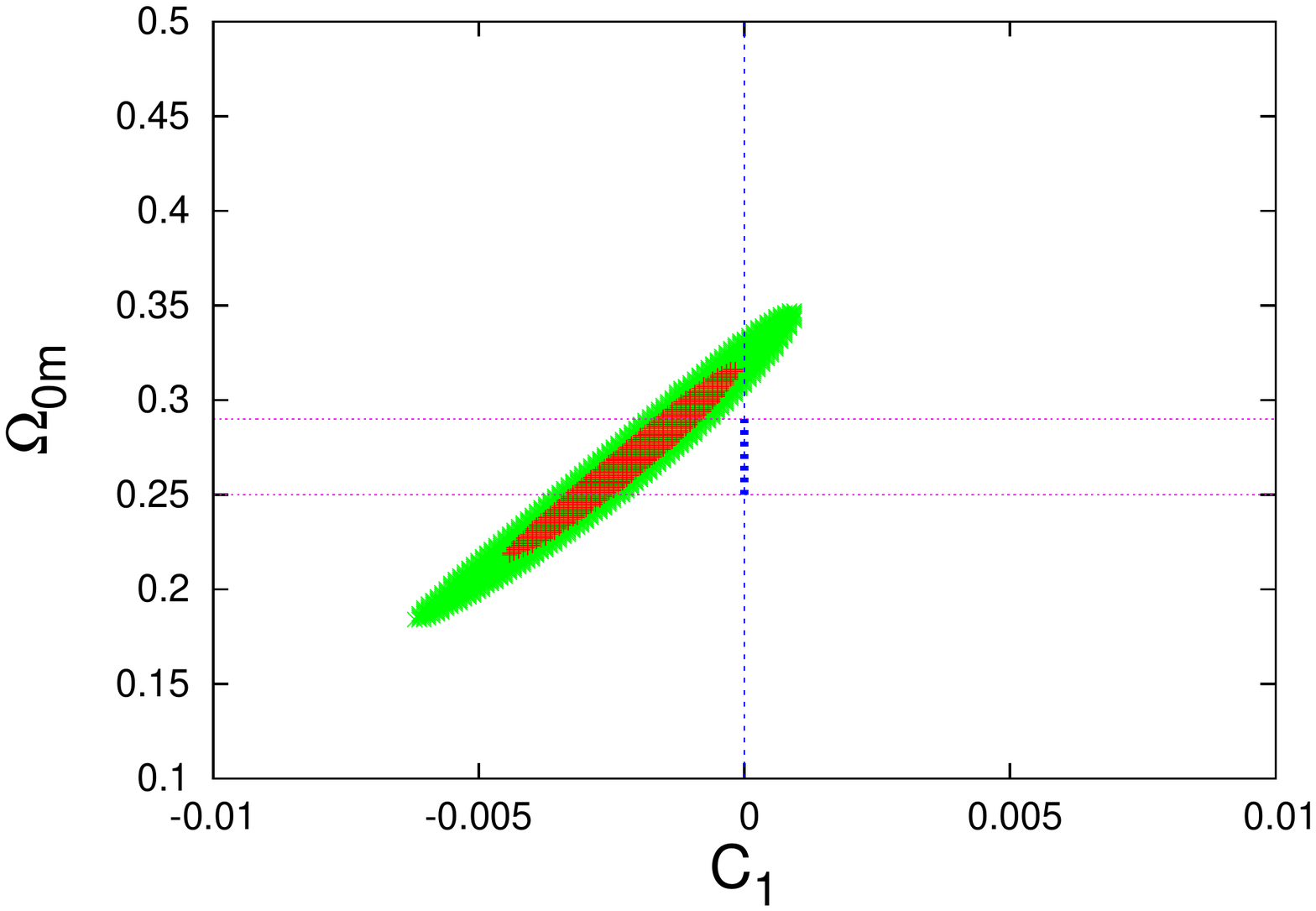}
\hspace{-1.in}
\includegraphics[scale=0.3, angle=0]{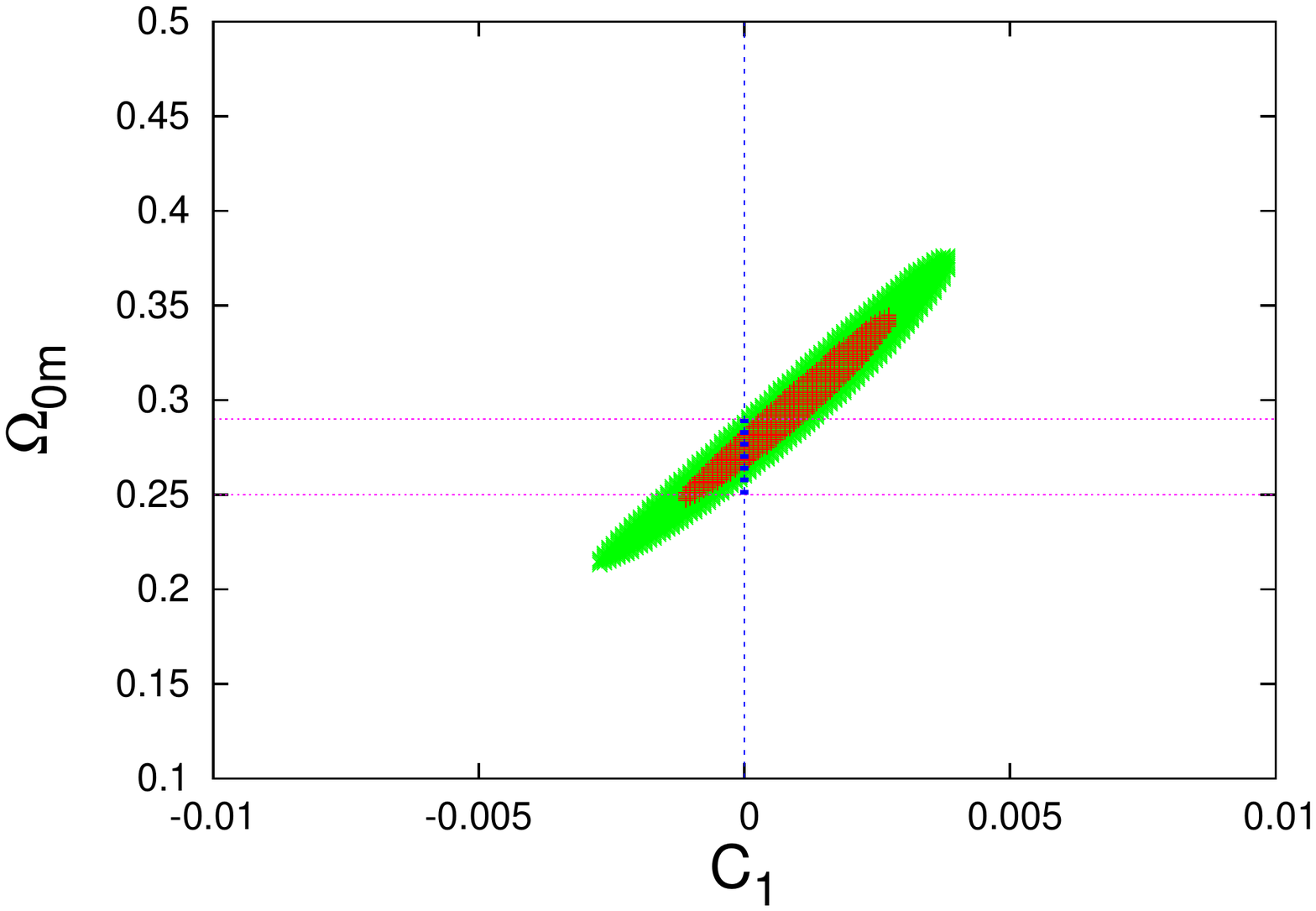}
\hspace{-1.in}
\includegraphics[scale=0.3, angle=0]{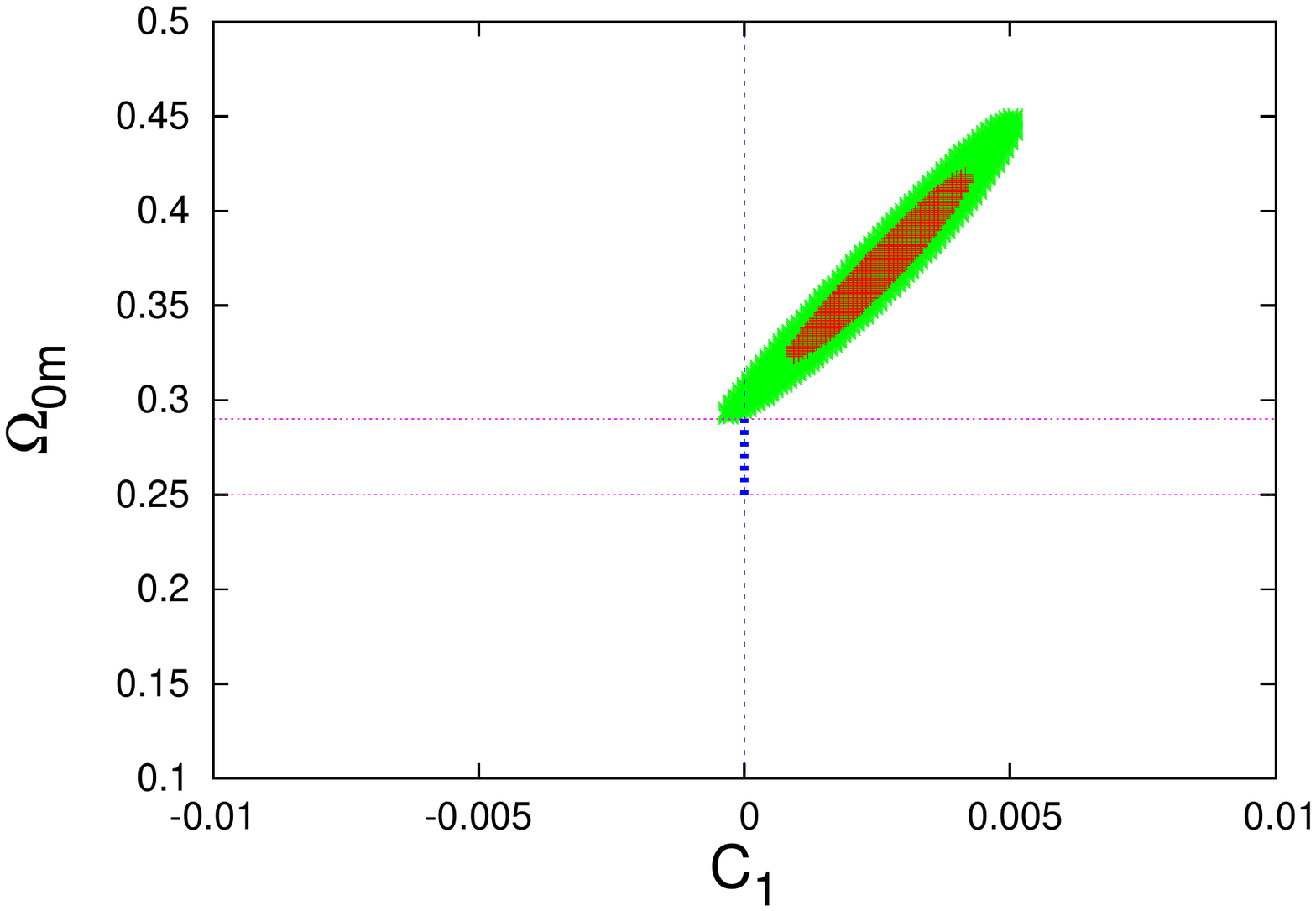}
\hspace{-1.1in}
\vspace{-0.2in}
\end{array}$
\vspace{-0.3in}
\end{center}
\caption {\small  $1\sigma$ and $2\sigma$ confidence limits of fitting $\mu_{DE}^{F_I}(z)$ assuming the theoretical model being $\Lambda$CDM to three different simulated data sets similar to figure 1 where in upper panels we have reduced the size of the error-bars for all data points by $20\%$ and in the lower panels we have increased the size of the error-bars for all data points by $20\%$. We can see that the method is robust against the uncertainties in the data up to a very high level.}  
\label{fig:int}
\end{figure}

To test the sensitivity of the method to the unknown dispersion of the data, $\sigma_{int}$, we re-do the analysis by increasing and decreasing the intrinsic dispersion of the data $\sigma_{int}$ by $20\%$ for all data points. In figure~\ref{fig:int} we see the results for fitting $\mu_{DE}^{F_I}(z)$ based on spatially flat $\Lambda$CDM model to the data with under and over estimated sizes of the error-bars. As we see, the method still works robustly and can differ between the models accurately even when we have large uncertainties in estimation of the intrinsic dispersion of the data.

The $w=-0.9$ model can be ruled out with high confidence and the kink model still shows inconsistency with the data at one to two sigma and having a knowledge of matter density from other cosmological observations we can rule out the kink model with high confidence.

\section{Conclusion}                        
\label{concl}

In summary, we have presented a new approach based on Crossing Statistics that can be used to distinguish between different cosmological models using Crossing functions and hyper-parameters and estimate the probability of an assumed model given the data. We propose that instead of parametrizing the cosmological quantities such as luminosity distance, Hubble parameter or equation of state of dark energy, we can parametrize the possible deviations from any assumed model and compare a model with its own variations. Performance of the method is shown being very much promising in distinguishing cosmological models with and without knowledge of matter density in comparison with other parametric and non parametric methods. Using this approach one can falsify the standard model (or any particular dark energy model) given the data without knowledge of the underlying actual model of the universe and its parameters. 
  
One may notice an interesting analogy between the first and second orders of Crossing functions in testing cosmological models using the cosmic distances and the spectral index and running of the spectral index of the primordial spectrum in CMB analysis. Similar to CMB analysis that we use the spectral index $n_s$ to judge about the consistency of the scale invariant power spectrum (Harrison-Zeldovich form of the primordial spectrum) to the data by looking at the derived $n_s$ and its error-bars (whether $n_s=1$ is in agreement to the CMB data), we use $C_1$ to test the consistency of the assumed dark energy model to the data (whether $C_1=0$ is in agreement to the distance data). Similarly there is an analogy between the running of the spectral index in CMB studies and $C_2$ in our analysis. One may argue that $n_s$ is a physical parameter while $C_1$ is not. We should note that independent of considering inflationary scenarios (or any physical mechanism resulting to a power-law form of the primordial spectrum) one can still use $n_s$ from a phenomenological perspective to rule out Harrison-Zeldovich form of the primordial spectrum.

The Bayesian interpretation of the Crossing Statistic appears to us to be a promising method of confronting cosmological models with supernovae observations, and has the potential to be trivially generalized to other cosmological observations.

\acknowledgments{AS thanks E. Linder, D. Pogosyan, L. Perivolaropoulos, T. Clifton, P. Ferreira and U. Seljak for useful discussions. AS thanks the anonymous referee for his/her valuable comments and suggestions. This work has been supported by World Class University grant R32-2009-000-10130-0 through the National Research Foundation, Ministry of Education, Science and Technology of Korea.}

\appendix

\end{document}